# Quantum nonlinear spectroscopy of single nuclear spins

Jonas Meinel[1*], Vadim Vorobyov[1*], Ping Wang[2,3*], Boris Yavkin[1], Mathias Pfender[1], Hitoshi Sumiya[4], Shinobu Onoda[5], Junichi Isoya[6], Ren-Bao Liu[2†], J. Wrachtrup[1,7 †]

1. 3rd Institute of Physics, Research Center SCoPE and IQST, University of Stuttgart, 70569 Stuttgart, Germany
2. Department of Physics, Centre for Quantum Coherence, and Hong Kong Institute of Quantum Information Science and Technology, The Chinese University of Hong Kong, Shatin, New Territories, Hong Kong, China
3. College of Education for the Future, Beijing Normal University at Zhuhai (BNU Zhuhai), Zhuhai, China
4. Sumitomo Electric Industries, Ltd., Itami 664-0016, Japan
5. Takasaki Advanced Radiation Research Institute, National Institutes for Quantum and Radiological Science and Technology, Takasaki 370-1292, Japan
6. Faculty of Pure and Applied Sciences, University of Tsukuba, Tsukuba 305-8573, Japan
7. Max Planck Institute for Solid State Research, Stuttgart, Germany

* These authors contributed equally

† Correspondence should be addressed to J.W. (wrachtrup@physik.uni-stuttgart.de) or R.B.L. (rbliu@cuhk.edu.hk).

**Nonlinear spectroscopy is widely used for studying physical systems [1]. Conventional nonlinear optical spectroscopy [1] and magnetic resonance spectroscopy [2,3], which use classical probes such as electromagnetic waves, can only access certain types of correlations in a quantum system [4]. The idea of quantum nonlinear spectroscopy [5] was recently proposed to use quantum probes such as entangled photons to achieve sensitivities and resolutions beyond the classical limits [6,7]. It is shown [8] that quantum sensing can extract arbitrary types and orders of correlations in a quantum system by first**

quantum-entangling a sensor and the object and then measuring the sensor [9,10]. Quantum sensing [11] has been applied to achieve nuclear magnetic resonance (NMR) of single atoms [12-14] and the second-order correlation spectroscopy [15-18] has been adopted to enhance the spectral resolution [18-21]. However, quantum nonlinear spectroscopy (i.e., the measurement of higher-order correlations) of single nuclear spins [8] is still elusive. Here we demonstrate the extraction of fourth-order correlations of single nuclear spins that cannot be measured in conventional nonlinear spectroscopy, using sequential weak measurement [9,20-25] via an atomic quantum sensor, namely, a nitrogen-vacancy center in diamond [26]. We show that the quantum nonlinear spectroscopy provides fingerprint features to identify different types of objects, such as Gaussian noises, random-phased AC fields, and quantum spins, which would be indistinguishable in second-order correlations. The measured fourth-order correlation unambiguously differentiates a single nuclear spin and a random-phased AC field. This work constitutes an initial step toward the application of higher-order correlations to quantum sensing [8,9], to examining the quantum foundation (by, e.g., higher-order Leggett-Garg inequality [27,28]), and to studying quantum many-body physics [29,30].

All information one can extract about a physical system is essentially statistics of measurement, quantified by correlations or moments. It is correlations that distinguish different types of noises or fluctuations. Higher-order correlations are particularly important since different types of physical quantities often have similar or only quantitatively different first- and second-order correlations [9,31,32]. For example, all the higher order correlations of Gaussian noises can be factorized into first- or second-order correlations of all possible partitions [33] and those of symmetric dichotomous telegraph noises can be factorized into second-order correlations only in sequential partitions [34].

Measuring correlations of fluctuations in physical systems is important to quantum science and technology. Correlations of measurements can test quantum foundations (such as Bell inequality [35] and Leggett-Garg inequality [27]) and identify the fundamental difference between classical and quantum systems [24,25]. The second-order correlation [15-18] detected by nitrogen-vacancy (NV) centers in diamond [26] has enabled high spectral resolution (1~100s Hz) in atomic NMR [18-21]. Quantum quantities, being operators, usually do not commute, i.e., two quantities $\hat{A}$ and $\hat{B}$ may have a non-zero commutator $[\hat{A}, \hat{B}] \equiv \hat{A}\hat{B} - \hat{B}\hat{A} \neq 0$, in sharp contrast to classical quantities, whose commutators always vanish. Therefore, quantum systems have characteristic quantum correlations, which involve commutators of quantities, such as the second-order one $\langle[\hat{A}, \hat{B}]\rangle$ (in which $\langle \cdots \rangle$ means average over many repeated measurements) and the third-order example $\langle\{\hat{A}, [\hat{B}, \hat{C}]\}\rangle$ (where $\{\hat{A}, \hat{B}\} \equiv \hat{A}\hat{B} + \hat{B}\hat{A}$ denotes an anti-commutator). Such quantum correlations can be used for classical-noise-free detection of quantum objects [4].

The rich structures of higher-order correlations are largely unexplored due to the limitation of conventional spectroscopy. In conventional nonlinear spectroscopy, a weak classical "force" $f_i$ is applied to a system at different times and/or locations, with a Hamiltonian $\hat{V}_i = f_i \hat{B}_i$, and the change of a physical quantity (the response) is measured. The response of a quantum system to the $K$-th order of the weak force is determined by a $(K+1)$-th order correlation that involves only commutator, such as $\langle[\hat{B}_1,[\hat{B}_2,\ldots[\hat{B}_K,\hat{A}]]]\rangle$; the response of a classical system contains only the classical correlation such as $\langle B_1 B_2 \cdots B_K A\rangle$. The correlations that involve anti-commutators, such as $\langle\{\hat{B}_1,[\hat{B}_2,\{\hat{B}_3,\hat{A}\}]\}\rangle$, do not show up in the response of a quantum system to a classical force. Similarly, the noise spectroscopy can also extract limited types of correlations [36-43].

Quantum probes in lieu of classical forces can be utilized to break the limits of conventional spectroscopy. Quantum light spectroscopy (using, e.g., entangled photons) has been demonstrated to have both high spectral and high temporal resolutions [5,7]. Quantum sensing provides a systematic approach to extracting higher-order correlations of arbitrary types [8]. A quantum sensor can establish entanglement with a quantum target, by which a measurement of the sensor constitutes a measurement of the target [9,10]. Specifically, one can perform a sequence of so-called weak measurements of a target by, repeatedly, weakly entangling the sensor with the target and measuring the sensor. By designing the initial state of the sensor and choosing the measurement basis in each shot of measurement, one can extract different types of correlations of the quantum target via statistics of the sequential outputs [8]. In conventional magnetic resonance spectroscopy, one can in principle separate the spin system into a quantum sensor and a target, but since the "sensor" is measured only at the end of a control sequence, the extractable correlations are restricted to those that can be coded by unitary quantum control or non-unitary ones that can be constructed from unitary controls via, e.g., phase cycling.

Here we demonstrate the extraction of fourth-order correlations of single nuclear spins using the statistics of sequential weak measurement. The measured correlations are inaccessible to conventional nonlinear spectroscopy. This first attempt of quantum nonlinear spectroscopy via quantum sensing already leads to non-trivial discoveries. We show that Gaussian noises, random-phased AC field, and single quantum spins exhibit qualitatively different patterns in the higher-order correlations. The measured fourth-order correlation unambiguously distinguishes a thermalized rotating nuclear spin and a classical AC field. It also provides a discrete count of the number of spins (similar to the photon-count correlation for determining the number of quantum emitters).

The sensing protocol is shown in Fig. 1 (a). In each shot of the sequential weak measurement, we prepare the sensor spin-1/2 in, e.g., the state $|x\rangle$. We then measure the sensor spin $\hat{\sigma}_\theta = \hat{\sigma}_x \cos\theta + \hat{\sigma}_y \sin\theta$ along the direction $\boldsymbol{e}_\theta$ (in the $xy$-plane with an angle $\theta$ from the $x$ axis). The weak interaction between the sensor and a quantum target $\hat{V}(t) = \hat{S}_z \hat{B}(t)$ (with $\hat{S}_z$ being the sensor spin along the $z$-axis and $\hat{B}(t)$ the quantum field from the target) can induce weak entanglement in an interrogation time $\tau$. The measurement on the sensor spin constitutes a weak measurement of the target. The correlations of the target can be extracted from the statistics of the measurement outputs $(\sigma_1, \sigma_2, \ldots, \sigma_j, \ldots)$ with $\sigma_j = \pm 1$. For example, the first moment $S_j = \langle \sigma_j \rangle$ was used to detect single nuclear spins [12-14], and the second moment $S_{ij} = \langle \delta\sigma_i \delta\sigma_j \rangle$ (with $\delta\sigma_i \equiv \sigma_i - \langle \sigma_i \rangle$) was measured for high-resolution atomic NMR [18-21]. Here we concentrate on the third moment $S_{ijk} = \langle \delta\sigma_i \delta\sigma_j \delta\sigma_k \rangle$. Not to be confused with the correlations in the targets, the $K$-th statistical moment of the measurement outputs will be referred to as the $K$-th order "signal".

Let us first consider a classical noise $B(t)$ along the $z$-axis. During the interrogation time $\tau$ in the $j$-th shot of measurement, the sensor spin precesses about the $z$-axis by an angle $\Phi_j \approx B_j \tau$ [where $B_j \equiv B(t_j)$]. The probability of output $\sigma_j = \pm 1$ of the measurement along $\boldsymbol{e}_\theta$ is $p_j(\pm) = \frac{1 \pm \cos(\theta - \Phi_j)}{2}$. For short interrogation time $\tau$ (in comparison to the timescale and the inverse strength of the noise), in the leading orders of coupling strength, the first moment is $S_j^C = \langle p_j \rangle \approx \cos\theta \left(1 - \langle \Phi_j^2 \rangle / 2\right)$

(where $p_j \equiv p_j^+ - p_j^-$), the second moment $S_{jk}^C = \langle p_j p_k \rangle \approx \sin^2\theta \langle \Phi_j \Phi_k \rangle$, and the third moment

$$S_{ijk}^C \approx -\frac{\sin^2\theta \cos\theta}{2}(\langle \delta\Phi_i^2 \Phi_j \Phi_k \rangle + \langle \Phi_i \delta\Phi_j^2 \Phi_k \rangle + \langle \Phi_i \Phi_j \delta\Phi_k^2 \rangle), \quad (1)$$

where $\delta\Phi_j^2 \equiv \Phi_j^2 - \langle \Phi_j^2 \rangle$. Here we have assumed that the noise is symmetric and therefore its odd-order correlations vanish. The phase correlations are related to the field correlations by $\langle \Phi_j \Phi_k \rangle = \tau^2 C_{jk}^C$ and $\langle \Phi_i \delta\Phi_j^2 \Phi_k \rangle = \tau^4 C_{ijjk}^C - \tau^4 C_{ik}^C C_{jj}^C$ with $C_{jk}^C \equiv \langle B_j B_k \rangle$ and $C_{ijkl}^C \equiv \langle B_i B_j B_k B_l \rangle$. The fourth-order correlations may be factorized into second-order ones with pairing patterns characteristic of the noise type. For example, a Gaussian noise allows all pairings, $C_{ijkl}^C = C_{ij}^C C_{kl}^C + C_{ik}^C C_{jl}^C + C_{il}^C C_{jk}^C$ [33], and an AC field with a uniformly random phase has $C_{ijkl}^C = (C_{ij}^C C_{kl}^C + C_{ik}^C C_{jl}^C + C_{il}^C C_{jk}^C)/2$ (the same as Gaussian noises, except for the factor 1/2) (see Supplementary Note 3). For a noise oscillating with angular frequency $\nu_0$, different types of statistics would yield the same second moment $S_{ij}^C \propto \cos(\nu_0 t_{ij})$ (with $t_{ij} \equiv t_i - t_j$). But the third moment $S_{ijk}^C$ (which contains the fourth-order correlation of the noise) would have different fingerprint patterns in its 2D spectrum $\tilde{S}^C(\nu_{ij}, \nu_{jk})$ (obtained by 2D Fourier transform in $t_{ij}$ and $t_{jk}$) for different types of noises (see Fig. 1b). In particular, the Gaussian noise has 12 peaks of equal height at $(0, \pm\nu_0)$, $(\pm\nu_0, 0)$, $\pm(\nu_0, \nu_0)$, $\pm(\nu_0, -\nu_0)$, $\pm(\nu_0, 2\nu_0)$, and $\pm(2\nu_0, \nu_0)$, and the random-phased AC field has six peaks at $\pm(2\nu_0, \nu_0)$, $\pm(\nu_0, -\nu_0)$, and $\pm(\nu_0, 2\nu_0)$ (see Supplementary Note 3).

The key difference between a quantum noise and a classical one is that in the interaction $\hat{V} = \hat{S}_z \hat{B}(t)$ the noise $\hat{B}(t)$ is an operator of the target (see Fig. 1a). In the $j$-th shot of measurement, starting from the initial state $\hat{\rho}(t_j) = \hat{\rho}_B(t_j) \otimes \hat{\rho}_S$ (where $\hat{\rho}_{B/S}(t_j)$ is the target/sensor state), the interaction leads to $\hat{\rho}(t_j + \tau) = \hat{\rho}(t_j) + \frac{\tau}{i}[\hat{V}(t_j), \hat{\rho}(t_j)] + \frac{\tau^2}{2i^2}[\hat{V}(t_j), [\hat{V}(t_j), \hat{\rho}(t_j)]] + \cdots$. To separate the effects on the sensor and those on the target, the commutator can be decomposed as $-i[\hat{V}(t_j), \hat{\rho}(t_j)] =$

$-i[\hat{S}_z, \hat{\rho}_S] \otimes \frac{1}{2}\{\hat{B}_j, \hat{\rho}_B(t_j)\} + \{\hat{S}_z, \hat{\rho}_S\} \otimes \frac{1}{2i}[\hat{B}_j, \hat{\rho}_B(t_j)] \equiv 2\mathbb{S}_z^- \hat{\rho}_S \otimes \mathbb{B}_j^+ \hat{\rho}_B + 2\mathbb{S}_z^+ \hat{\rho}_S \otimes \mathbb{B}_j^+ \hat{\rho}_B$, where $\mathbb{B}^+ \hat{A} \equiv (\hat{B}\hat{A} + \hat{A}\hat{B})/2$ (essentially the anti-commutator) reduces to the normal product if $\hat{B}(t)$ is a classical field and $\mathbb{B}^- \hat{A} \equiv (\hat{B}\hat{A} - \hat{A}\hat{B})/(2i)$ (essentially the commutator) vanishes if $\hat{B}(t)$ is a classical field. By choosing to measure the sensor in the basis of $\mathbb{S}_z^+ \hat{\rho}_S$ or $\mathbb{S}_z^- \hat{\rho}_S$, one can select the target evolution driven by the commutator or the anti-commutator, i.e., $\mathbb{B}_z^- \hat{\rho}_S$ or $\mathbb{B}_z^+ \hat{\rho}_S$, respectively. Therefore, quantum correlations that contain a nested sequence of commutators and anti-commutators of the noise operators can be extracted. Considering a target (such as a nuclear spin) at high temperature, i.e., $\hat{\rho}_B$ being a constant, the second order quantum correlation $\text{Tr}(\mathbb{B}_j^+ \mathbb{B}_i^- \hat{\rho}_B)$ vanishes. The third momentum has both classical and quantum contributions, $S_{ijk} = S_{ijk}^C + S_{ijk}^Q$.

The classical part $S_{ijk}^C$ is the same as for classical noises (see Methods and Supplementary Note4), except that the products of classical variables should be replaced with anti-commutators such as ($t_l > t_k > t_j > t_i$)

$$C_{ijkl}^C = \text{Tr}(\mathbb{B}_l^+ \mathbb{B}_k^+ \mathbb{B}_j^+ \mathbb{B}_i^+ \hat{\rho}_B). \tag{2}$$

It should be noted that though the classical correlation in the quantum object takes the same form as in a classical noise, it has a fundamentally different origin. The correlations in the quantum object stem from the back-action of the weak measurement by the sensor, which results from the weak entanglement and measurement of the sensor in the basis of $\mathbb{S}_z^- \hat{\rho}_S$. Importantly, the classical correlation $C_{ijjk}^C$ of a *quantum* object in Eq. (2) does not contribute to conventional nonlinear spectroscopy using a *classical* probe.

The quantum part $S_{ijk}^Q = -\frac{1}{2}\sin^2\theta \cos\theta \, \tau^4 C_{ijjk}^Q$ (see Mehtods). For $\hat{\rho}_B$ being a constant, the quantum correlation has only one non-vanishing term ($t_k > t_j > t_i$ assumed, see Methods for details)

$$C_{ijjk}^Q = \text{Tr}(\mathbb{B}_k^+ \mathbb{B}_j^- \mathbb{B}_j^- \mathbb{B}_i^+ \hat{\rho}_B). \tag{3}$$

The importance of quantumness lies in the fact that without the heralded polarization of the target by back-action from measurement at $t_i$, the commutators at $t_j$ would vanish [4,8].

When the quantum object is a two-level system (such as spin-1/2 of $^{13}$C in diamond), the quantum correlations will double the third moment since $S_{ijk}^Q = S_{ijk}^C$ in this case (see Methods). When the sensor is coupled to multiple ($N$) spin-1/2's at high temperature (see Methods), the classical correlation scales as $C_{ijjk}^C \sim N^2$, and the quantum correlation $C_{ijjk}^Q \sim N$ (since the commutators between different spins vanish). With increasing the number, the quantum spins approach to a classical noise, with a Gaussian statistics (resulting from the summation of many independent binary quantities). Figure 1c shows qualitatively different patterns in the correlation spectra of different number of spin-1/2's with uniform coupling.

We employ the states $|+\rangle = |0_e\rangle$ and $|-\rangle \equiv |-1_e\rangle$ in the spin triplet of an NV center in diamond as the sensor spin [26]. Each shot of weak measurement is realized by the pulse sequence shown in Fig. 2a. We optically pump the NV center spin into the state $|+\rangle$ and prepare it into the state $|x\rangle = (|+\rangle + |-\rangle)/\sqrt{2}$ by a $\frac{\pi}{2}$ microwave pulse. A sequence of dynamical decoupling pulses modulates the interaction between the NV spin and a target $^{13}$C nuclear spin during the interrogation such that weak, tunable entanglement between the sensor and the target is induced. Measurement of $\hat{\sigma}_\theta$ is realized by a $\frac{\pi}{2}$ rotation changing the $\boldsymbol{e}_\theta$ axis to the $z$ axis followed by a projective measurement along the $z$ axis. To enhance the readout fidelity, we use a SWAP gate to store the sensor spin state in the $^{14}$N nuclear spin (which has been polarized in the initialization step using SWAP gates as well) and repeatedly ($M$ times) read out the $^{14}$N spin via a CNOT gate and spin-dependent fluorescence of the NV center electron spin [44,45]. The statistical moments of the measurement $S_i$, $S_{ij}$, and $S_{ijk}$ are reconstructed from the photon counts (see Methods).

Figure 2b shows the second-order signal $S_{ij}$ of a sensor coupled to a $^{13}$C nuclear spin. Under an external magnetic field ($B_0$ =0.2502 T) along the z direction (the NV axis) and dynamical decoupling control of the hyperfine interaction, the quantum field from the $^{13}$C spin in the interaction picture is effectively $\hat{B}(t) = A_\perp[\hat{I}_x \cos(\nu_0 t) - \hat{I}_y \sin(\nu_0 t)]$ (see Methods) with the nuclear Zeeman frequency $\frac{\nu_0}{2\pi} \approx$ 2.6795 MHz . Therefore, $S_{ij} \propto C_{ij}^C = \frac{1}{2}\langle\{\hat{B}_i, \hat{B}_j\}\rangle \propto \cos(\nu_0 t_{ij}) e^{-\gamma t_{ij}}$, oscillates at frequency $\nu_0$ with a measurement-induced decay [9] (a rapid decay due to random hopping of the NV center state has been removed – see Supplementary Note 9). For comparison, Fig. 2c shows both the Fourier transform of the second-order signal for an AC field $B(t) = B_0 \cos(\nu_0 t + \phi)$ with a uniformly random phase $\phi$ and that for a $^{13}$C nuclear spin. As shown in Fig. 2c, the nuclear spin and the random-phased AC field lead to similar second-order signals.

The third moment of the sequential measurements has qualitatively different patterns for a quantum spin and for a classical field. We set the measurement angle $\theta \approx$ 54.7° to maximize the amplitude of the third-order signal (which is $\propto \sin^2\theta \cos\theta$). The 2D spectrum of the third moment for a quantum spin target (Figs. 3a & 3b) clearly shows four peaks at $(\nu_{ij}, \nu_{jk})$ with $|\nu_{ij}| = |\nu_{jk}| = \nu_0$ mod $(2\pi/t_c)$ with $t_c$ being the period of each measurement shot. The difference in the heights of the diagonal and anti-diagonal peaks is probably due to the fast hopping between different states of sensor spin (see Supplementary Note 6). In contrast, the 2D spectrum for the random-phased AC field (Fig. 3c), as expected, presents six peaks at $(\nu_{ij}, \nu_{jk}) = \pm(\nu_0, 2\nu_0)$, $\pm(\nu_0, -\nu_0)$, and $\pm(2\nu_0, \nu_0)$ mod $(2\pi/t_c)$.

The quantum nonlinear spectroscopy has qualitatively different patterns for different number of nuclear spins (Fig. 1c). In particular, the height of the eight peaks at $(0, \pm\nu_0), (\pm\nu_0, 0), \pm(\nu_0, 2\nu_0),$ or $\pm(2\nu_0, \nu_0)$ relative to those at $\pm(\nu_0, -\nu_0)$ is a quantized number $\eta = 1 - 1/N$ (see Fig. 3d and Supplementary Note 5), which provides a discrete count of the number of spins (similar to the determination of the number of quantum emitters by the correlation $g^{(2)}$ of photon counts). The relative height $\eta$ averaged over the signals at the eight points is about $0.12 \pm 0.1$(Fig. 3d), indicating that the target detected by the sensor is a single nuclear spin. Instead of roughly estimating the number of nuclear spins by sensitivity [46], our method can determine the exact number if the couplings to different spins are of similar strength.

The third moment contains the contribution of the quantum correlation and hence can differentiate a quantum spin and a classical noise. In particular, the second moments for a spin-1/2 at higher temperature is $S_{jk} = c_0 \sin^2\theta \cos(v_0 t_{jk}) e^{-\gamma t_{jk}}$ (with $c_0$ being a constant), the third-order signal for a quantum spin target is $S_{ijk} = -rc_0^2 \sin^2\theta \cos\theta \sin(v_0 t_{ij}) \sin(v_0 t_{jk}) e^{-\gamma t_{ik}}$ (see Methods), with $r = 1$ for a quantum spin target and $r_c = 1/2$ for the classical signal $S_{ijk}^C$. The fitted result, as shown in Fig. 4d, yield $r = 1.13$ with a standard deviation $\approx 0.368$. The data confirms the quantumness of the noise from the nuclear spin.

These results demonstrate that quantum nonlinear spectroscopy, enabled by measurement mediated by a quantum sensor, can extract correlations of a quantum object that are inaccessible to conventional nonlinear spectroscopy using classical probes and quantum correlations that are missing in classical fields and cannot be retrieved by conventional noise spectroscopy. The higher-order correlations provide fingerprint features for unambiguous differentiation of noises of different nature and for verification of quantumness. The scheme can be generalized, by using, e.g., different initial states of the sensor spin, different measurement bases, higher orders moments, and higher spins as sensors, to extract arbitrary types and orders of correlations. Information made available by quantum nonlinear spectroscopy will be useful for quantum computing (by helping characterize and optimally suppress noises), quantum sensing (by isolating quantum objects from classical noise background), studying quantum many-body physics (by detecting new types of fluctuations in mesoscopic systems), and examining quantum foundation (by testing higher-order Bell inequalities or Leggett-Garg inequalities with fewer or no interpretation loopholes).

## Methods

### Setup and sample

The measurement is carried out with a confocal microscope setup located in a room temperature bore of a superconducting magnet (see Supplementary Fig. S1). The magnet produces a field of 250 mT, aligned parallel to the NV axis, which results in a transition frequency of about 4.1 GHz between $|0\rangle$ and $|-1\rangle$. The fluorescence light of the NV centers are detected with an avalanche photo diode (APD). The electron and nuclear spins are manipulated with the two channels of microwaves. We have a typical Rabi frequency of 7 MHz for the electron spin at full pulse amplitude.

The diamond sample used is a 2 mm × 2 mm × 80 μm, (111)-oriented polished slice from a $^{12}$C-enriched (99.995%) diamond crystal [20]. The single NV centers were created by electron irradiation. The typical lifetimes for the NV centers in this slice are $T_2^* \approx 50$ μs (measured by Ramsey interference) and $T_2 \approx 300$ μs (measured by spin echo).

For details of the setup and the sample see Supplementary Note 1.

### Measurement method

We use the NV center electron spin as the sensor and the nitrogen nuclear spin as a quantum memory to enhance the sensing. Each shot of measurement consists of three steps: initialization, sensing and readout. The electron spin is optically initialized in state $|0\rangle$. The electron spin is prepared with a $(\pi/2)$-pulse. We sense $^{13}$C nuclear spins in diamond with a Knill pulse dynamical decoupling sequence [47], the KDD$xy$, where the time between pulses matches the Larmor frequency of the $^{13}$C spin of interest. Therefore, the superposition state of the NV electron spin acquires a phase, conditioned on the $^{13}$C state. The sensor state gets projected, orthogonal to the preparation pulse, with a phase-shifted MW pulse. Lastly the optical readout of the sensor spin is performed. Here we SWAP the electron spin state and the $^{14}$N spin state, which is preserved during several laser readouts, enabling single-shot readout [44,45]. To mitigate the effect of decoherence because of the hyperfine interaction with the target $^{13}$C when the NV center is in the excited states, we limit the readout to 40 repetitions. The SWAP between the electron and the memory spins consists of a weak MW pulse on the electron spin conditional on the $^{14}$N state (CNOTe, with duration ~4 μs) followed by a conditional RF-pulse on the memory spin (CNOTn, with duration ~50 μs) and then another CNOTe. Each readout repetition consists of one CNOTe and a laser pulse (0.3 μs).

**Reconstruction of correlation from photon counts**

The probability for sensor collapse to $|0\rangle$ ($|-1\rangle$) is denoted by $p(+)$ ($p(-)$). For weak noise, $p(\sigma) \approx [1 + \sigma \cos(\theta - \Phi)]/2$. The distribution of photon counts of each measurement is

$$p(n) = p(n|+)p(+) + p(n|-)p(-),$$

where $p(n|\pm) = \frac{1}{n!} e^{-n_\pm} n_\pm^n$ is the Poisson distribution and $n_\pm$ is the average number of photons detected for the spin state $|0\rangle$ or $|-1\rangle$, respectively. The photon counts can be written as

$$n = \bar{n} + \sigma d + w_\sigma,$$

with $\bar{n} = (n_+ + n_-)/2$ is the average photon count, $d \equiv (n_+ - n_-)/2$ is the photon count contrast between the two spin states, and $w_\sigma \equiv n - n_\sigma$ is the intrinsic photon count fluctuation (due to spontaneous emission, APD efficiency, etc) satisfying the distribution $p(w_\sigma) = p(n_\sigma + w_\sigma|\sigma)$ with zero mean value. The photon count fluctuation $\delta n_i \equiv n_i - \langle n_i \rangle$ is related to the spin signal fluctuation $\delta \sigma_i \equiv \sigma_i - \langle \sigma_i \rangle$ by

$$\delta n_i = \delta \sigma_i d + w_{\sigma_i},$$

with the first moment of the photon counts being $\langle n_i \rangle = \bar{n} + \langle \sigma_i \rangle d \approx \bar{n} + c \cos \theta$. The second and third moments are respectively

$$\langle \delta n_j \delta n_i \rangle = d^2 \langle \delta \sigma_j \delta \sigma_i \rangle,$$

$$\langle \delta n_k \delta n_j \delta n_i \rangle = d^3 \langle \delta \sigma_k \delta \sigma_j \delta \sigma_i \rangle,$$

for $i, j, k$ being different. Here we have used the fact that the intrinsic photon count fluctuations $w_{\sigma_j}$ are independent for different shots of measurements.

**Effective Hamiltonian under dynamical decoupling**

The evolution during the interrogation in the interaction picture is $\hat{U} = \hat{T} \exp\left(-i \int_t^{t+\tau} f(u) \hat{V}_{\text{hf}}(u) du\right) = \exp(-i \hat{V} \tau)$, where $f(u)$ is the modulation function alternating between $+1$ and $-1$ due to the dynamical decoupling sequence [33] and $\hat{V}_{hf}(u)$ is the hyperfine interaction in the interaction picture. By Magnus expansion for short period of time, the effective coupling $\hat{V}(t) \approx \tau^{-1} \int_{t}^{t+\tau} f(u) \hat{V}_{hf}(u) du$. For the coupling to a single $^{13}$C spin, $\hat{V}_{hf}(t) = A_x \hat{S}_z \hat{I}_x(t)$ with $\hat{I}_x(t) = \hat{I}_x \cos(\nu_0 t) - \hat{I}_y \sin(\nu_0 t)$. Under the KDD, the effective coupling becomes $\hat{V}(t) \approx A_\perp \hat{S}_z \hat{I}_x(t)$ [20], with $A_\perp = 2A_x/\pi$.

**Quantum correlations**

The relation between the statistics (moments) of the sequential measurement and the correlation of the noise field $\hat{B}(t)$ can be directly obtained by the perturbative expansion of the evolution during interrogation time $\tau$. We assume that the bath evolves freely between two adjacent interrogation processes. As shown in Fig. 1a, the measurements at different times, though conducted on a single NV center spin in the experiment, can be viewed as performed independently on different sensor spins $\{\hat{S}_j\}$, each interacting with the target with Hamiltonian $\hat{V}_j = \hat{S}_{j,z} \hat{B}_j$ from $t_j$ to $t_j + \tau$. The initial state of the target and the sensors can be written as $\hat{\rho} = \hat{\rho}_B \otimes \hat{\rho}_1 \otimes \hat{\rho}_2 \otimes \cdots$ with $\hat{\rho}_j = |x\rangle\langle x| = \hat{S}_{j,x} + \frac{1}{2}$ for the $j$-th sensor spin and $\hat{\rho}_B = 2^{-N}$ being the density operator of $N$ nuclear spins at high temperature. The evolution due to the interaction with the $j$-th sensor can be expanded as

$$\hat{\rho}(\tau) = \hat{\rho} + \frac{\tau}{i}[\hat{V}_j, \hat{\rho}] + \frac{1}{2!}\left(\frac{\tau}{i}\right)^2 [\hat{V}_j, [\hat{V}_j, \hat{\rho}]] + \cdots$$

The first moment of the measurement is

$$S_j = \langle \hat{\sigma}_{j,\theta} \rangle = \cos\theta - \frac{1}{2!}\tau^2 \text{Tr}_S\left(\hat{\sigma}_{j,\theta}\left[\hat{S}_{j,z}, [\hat{S}_{j,z}, \hat{\rho}_j]\right]\right) \text{Tr}_B(\mathbb{B}_j^+ \mathbb{B}_j^+ \hat{\rho}_B) + \cdots$$
$$= \cos\theta \left(1 - \frac{1}{2}\tau^2 C_{jj}^C + \cdots\right).$$

The second moment (for $t_j > t_k$) is $S_{jk} = \langle \delta\hat{\sigma}_{j,\theta} \delta\hat{\sigma}_{k,\theta} \rangle$, where $\delta\hat{\sigma}_{j,\theta} \equiv \hat{\sigma}_{j,\theta} - \langle \hat{\sigma}_{j,\theta} \rangle$. Since in the zeroth order of the fluctuation $\langle \delta\hat{\sigma}_{j,\theta} \rangle = 0$, the second moment must contain at least one order of the noise field at each time. Thus, in the leading order of the noise field, the second moment is

$$S_{jk} \approx \frac{\tau^2}{i^2} \text{Tr}\big(\delta\hat{\sigma}_{j,\theta}[\hat{S}_{j,z},\hat{\rho}_j]\big)\text{Tr}\big(\delta\hat{\sigma}_{k,\theta}[\hat{S}_{k,z},\hat{\rho}_s]\big)\text{Tr}\big(\text{B}_j^+\text{B}_k^+\hat{\rho}_B\big) = \tau^2 \sin^2\theta\, C_{jk}^C.$$

The third moment $S_{ijk} = \langle\delta\hat{\sigma}_{i,\theta}\delta\hat{\sigma}_{j,\theta}\delta\hat{\sigma}_{k,\theta}\rangle$ can be similarly obtained as

$$S_{ijk} \approx -\frac{\tau^4 \cos\theta \sin^2\theta}{2}\big(C_{iijk}^C - C_{ii}^C C_{jk}^C + C_{ijjk}^C - C_{ik}^C C_{jj}^C + C_{ijkk}^C - C_{ij}^C C_{kk}^C + C_{ijjk}^Q\big).$$

See Supplementary Note 4 for details.

**Correlations of *N* uniformly coupled nuclear spins**

The noise field from $N$ uniformly coupled nuclear spins $\{\hat{I}_n\}$ can be written as $\hat{B}(t) = \sum_{n=1}^{N} A_\perp[\hat{I}_{n,x}\cos(\nu_0 t) + \hat{I}_{n,y}\sin(\nu_0 t)]$ in the interaction picture. With $\hat{\rho}_B = 2^{-N}$, the second-order classical correlation is

$$C_{jk}^C = \text{Tr}[\text{B}_j^+\text{B}_k^+\hat{\rho}_B] = \frac{1}{4}NA_\perp^2 \cos(\nu_0 t_{ij}) e^{-\gamma t_{jk}} \sim O(N),$$

where decoherence of nuclear spins (due to, e.g., back-actin of the weak measurement between $t_j$ and $t_k$) is taken into account as the exponential decay factor (see Ref. [20]). The fourth order classical correlation is

$$C_{ijkl}^C \equiv \text{Tr}\big(\text{B}_i^+\text{B}_j^+\text{B}_k^+\text{B}_l^+\hat{\rho}_B\big) = C_{ij}^C C_{kl}^C + \frac{N-1}{N}\big(C_{ik}^C C_{jl}^C + C_{il}^C C_{jk}^C\big)\sim O(N^2),$$

which is the same as for a telegraph noise for $N = 1$ and approaches to the Gaussian noise for $N \gg 1$. The second order quantum correlation $C_{ij}^Q = 0$ and the fourth order

$$C_{ijkl}^Q \equiv \text{Tr}\big(\text{B}_i^+\text{B}_j^-\text{B}_k^-\text{B}_l^+\hat{\rho}_B\big) = \frac{1}{16}NA_\perp^4 \sin(\nu_0 t_{ij})\sin(\nu_0 t_{kl}) e^{-\gamma t_{ij}-\gamma t_{kl}} \propto N,$$

which is much smaller than the classical correlation for $N \gg 1$. For $N = 1$, the classical contribution $S_{ijk}^C \propto C_{ijjk}^C - C_{jj}^C C_{ik}^C$ and the quantum one $S_{ijk}^Q \propto C_{ijjk}^Q$ are equal, and the total third moment

$$S_{ijk} = 2S_{ijk}^C \propto \sin(\nu_0 t_{ij})\sin(\nu_0 t_{jk}) e^{-\gamma t_{ik}}.$$

Its 2D Fourier transform $\tilde{S}(\nu_{ij}, \nu_{jk})$ has four peaks at $\pm(\nu_0, \nu_0)$ and $\pm(\nu_0, -\nu_0)$, with equal height.

**Acknowledgements**

VV, JM, MP, JW acknowledge financial support by the German Science Foundation (the DFG) via SPP1601, FOR2724, the European Research Council (ASTERIQS, SMel, ERC grant 742610), the Max Planck Society, and the project QC4BW as well as QTBW. P.W. & R.B.L are supported by Hong Kong Research Grants Council General Research Fund Project 143000119.

**Author Contributions**

R.B.L. & J.W. conceived the project. P.W. & R.B.L. carried out the theoretical study; J.W. supervised the experimental studies. V.V., J. M., B.Y. & M.P. carried out the measurements; H.S., H.O. & J.I. designed and conducted the synthesis and fabrication of the diamond substrate; V.V., J.M., P.W., R.B.L. & J.W. analyzed the data; R.B.L., P.W., V.V. & J.W. wrote the manuscript with contributions from other authors.

**Competing financial interests**

The authors declare no competing financial interests.

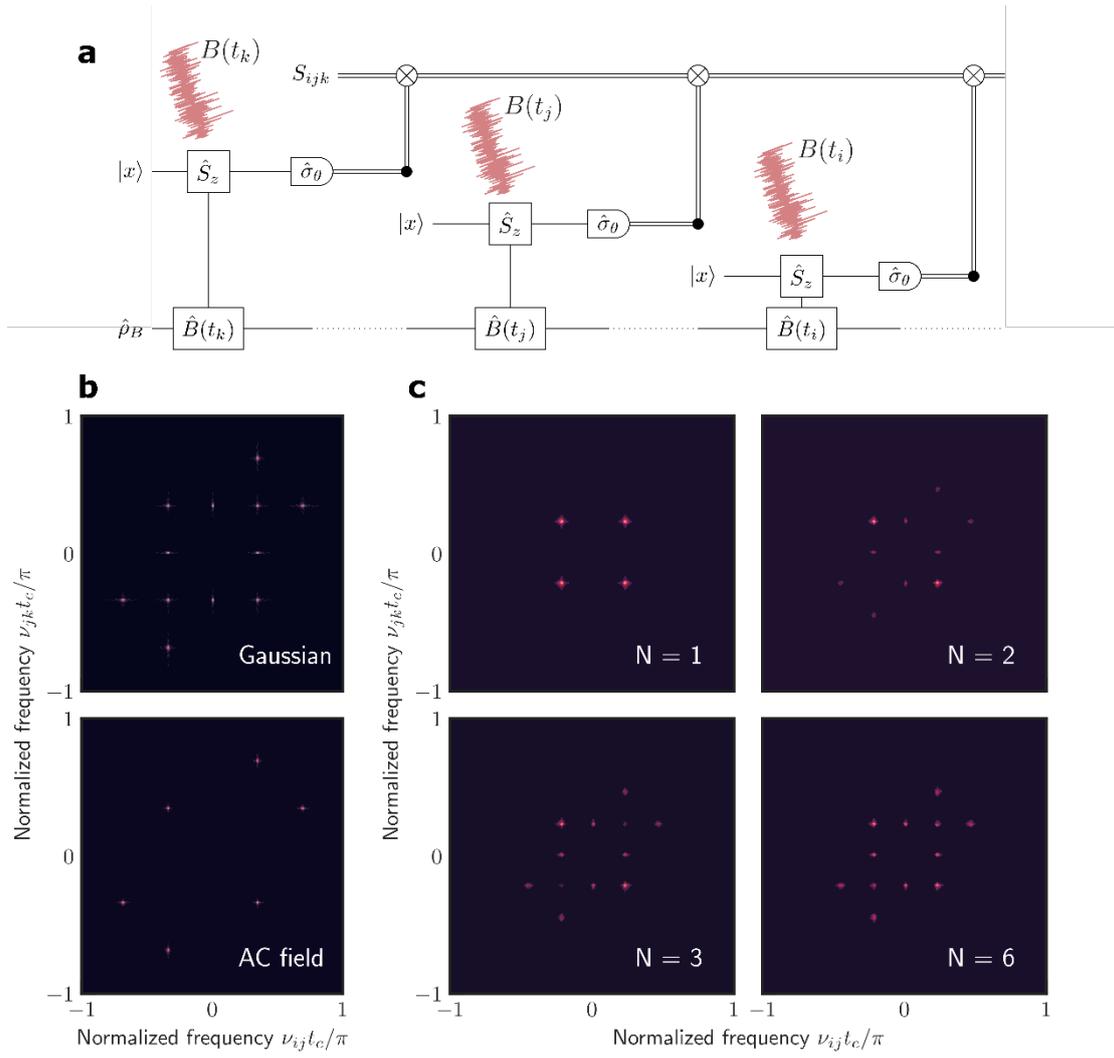

**Fig. 1. Fourth-order correlation spectra of different types of classical and quantum noises. a.** The scheme of correlation measurement. A sensor spin is initially prepared in the state $|x\rangle$, then its $z$-component $\hat{S}_z$ is coupled to a classical noise $B(t)$ or a quantum object by $\hat{B}(t)$, and the measurements along $\boldsymbol{e}_\theta$ are correlated to determine the statistical moments, e.g., $S_{ijk}$. **b.** 2D spectra $\tilde{S}(\nu_{ij}, \nu_{jk})$ of the third moment $S_{ijk}$ for a Gaussian noise (upper) and a random-phased AC field (lower). **c.** 2D spectra of the third moment for a sensor coupled uniformly to $N$ spin-1/2's ($N = 1, 2, 3,$ and $6$).

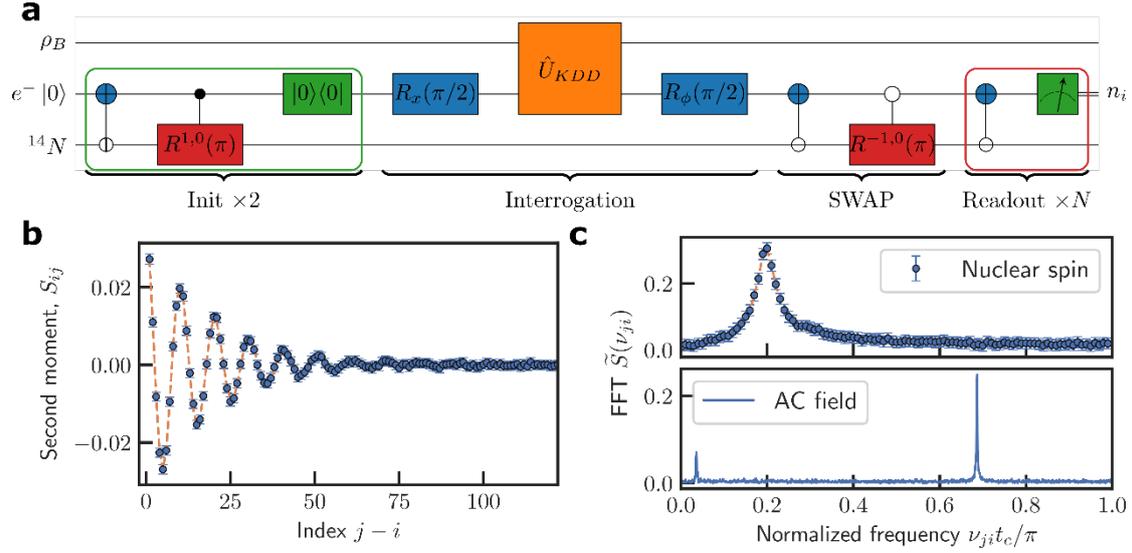

**Fig. 2. Statistics of sequential measurements on a sensor spin. a.** Protocol of sequential measurement. The sensor spin and the ancilla are initialized by optical pump (green block being a pulse of 532 nm laser) and SWAP gates (repeated twice for higher fidelity). Then the sensor spin is rotated by a $\frac{\pi}{2}$ pulse (blue block), controlled by a dynamical decoupling sequence, and rotated again by a $\frac{\pi}{2}$ pulse (with a readout angle $\theta$ from the first $\frac{\pi}{2}$ pulse so that $\hat{\sigma}_\theta$ is measured). The NV electron spin state is then stored in the $^{14}$N spin by a SWAP gate and the $^{14}$N spin state is repeatedly read out through the electron spin via a CNOT gate and photon counts. **b.** Second moment $S_{ij}$ of sequential measurement of a sensor spin coupled to a $^{13}$C nuclear spin. **c.** Fourier transform of the second moment for a nuclear spin (upper) and a random-phased AC field (lower). The extra small peak at lower frequency in lower graph of (**c**) is not from the AC field, as checked by the dependence of its amplitude on the measurement direction $\boldsymbol{e}_\theta$ (see Figure S5 in Supplementary Note 9).

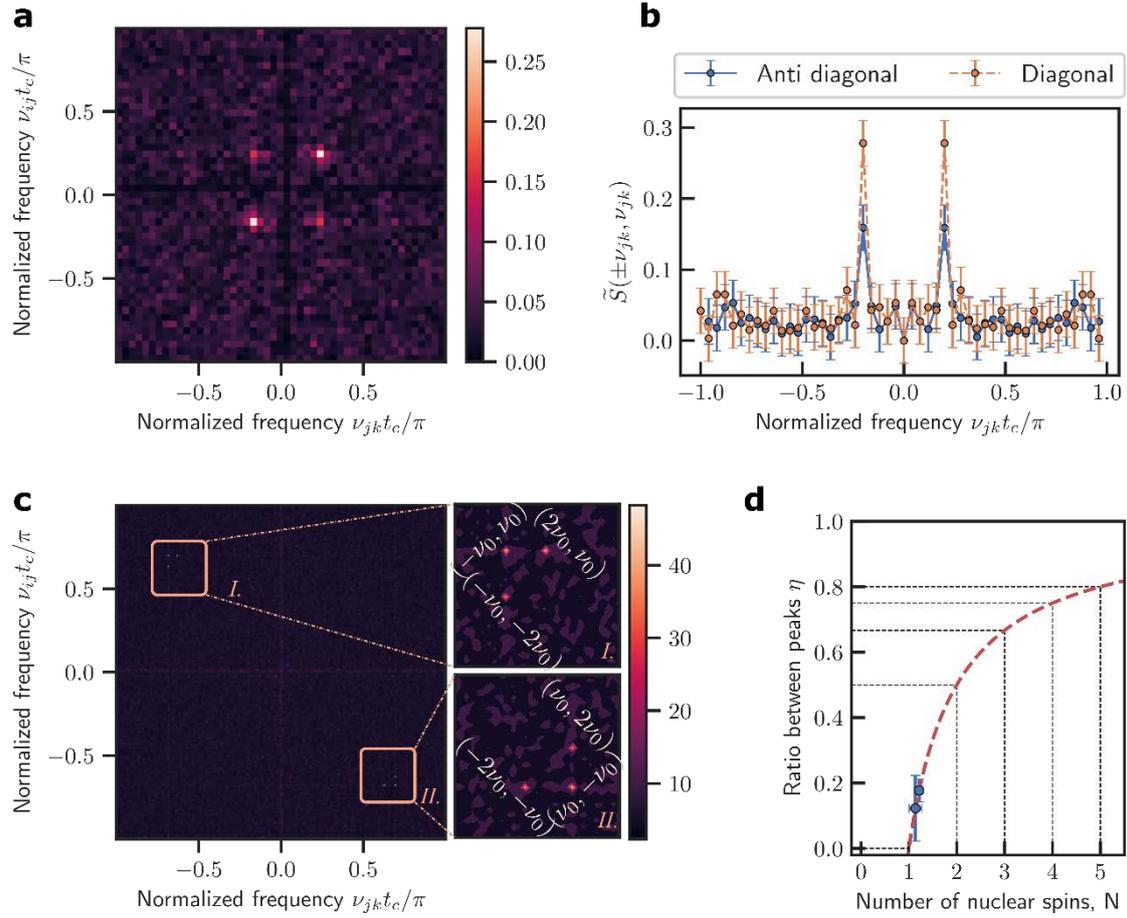

**Fig. 3. Quantum nonlinear spectroscopy of a nuclear spin and a random-phased AC field. a.** 2D spectrum of the third moment $\tilde{S}(\nu_{ij}, \nu_{jk})$ of an NV center spin coupled to a nuclear spin. **b.** The diagonal (orange symbols) and anti-diagonal (purple symbols) slice of (**a**). **c.** 2D spectrum of the third moment $\tilde{S}(\nu_{ij}, \nu_{jk})$ of an NV center spin coupled to a random-phased AC field. **d.** The calculated average height (curve) of the eight peaks at $(0, \pm\nu_0)$, $(\pm\nu_0, 0)$, $\pm(\nu_0, 2\nu_0)$, or $\pm(2\nu_0, \nu_0)$ relative to those at $\pm(\nu_0, -\nu_0)$ as a function of the number of uniformly coupled nuclear spins. The symbols are experimental values (green one from Fig. 3a and blue one from Supplementary Figure S12, measured with a different number of dynamical decoupling pulses). Error bars are standard deviation.

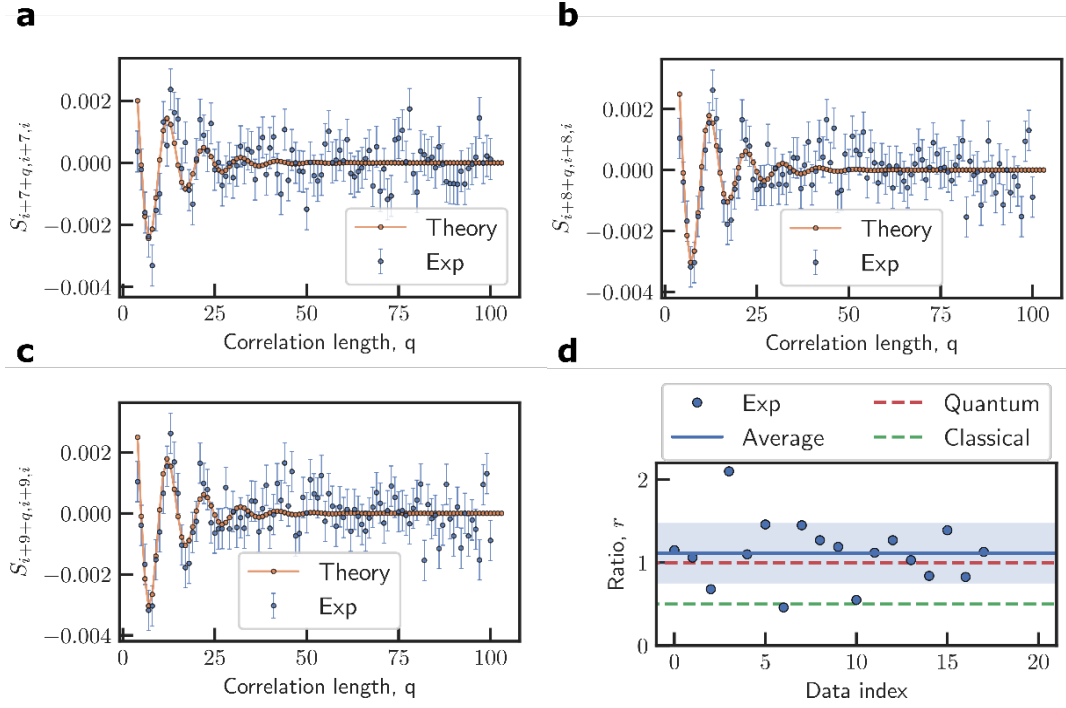

**Fig. 4. Quantum correlation of a single nuclear spin. a**, **b** and **c** show the third moment $S_{i,i+p,i+p+q}$ as a function of $p$ for $q = 7, 8$, and $9$ in turn. The purple symbols are experimental data. The orange curves are theoretical results with the fitting parameter $r$ being the ratio of the amplitude of the third moment to the amplitude squared of the second moment (not shown). **d.** The factor $r$ (purple symbols) obtained from fitting different data sets (see Supplementary Note 11). The purple line is the mean value of $r$, and the shadow area is within one standard deviation from the mean. The blue (yellow) dashed line indicates the value $r_Q = 1$ ($r_C = 1/2$) for the total (classical only) correlations.